\documentclass[aps,pra,reprint,10pt,raggedbottom,superscriptaddress,nofootinbib]{revtex4-2}
\usepackage{amsmath,amssymb,amsthm}
\usepackage{microtype}
\usepackage[colorlinks=true,linkcolor=blue,citecolor=blue,urlcolor=blue]{hyperref}
\hypersetup{pdftitle={Critical properties of EPR steering},pdfauthor={Yu-Xuan Zhang and Jing-Ling Chen}}
\makeatletter
\let\auto@bib@innerbib\@empty
\makeatother

\AtBeginDocument{}
\newtheorem{theorem}{Theorem}
\newtheorem{corollary}{Corollary}
\newtheorem{proposition}{Proposition}
\DeclareMathOperator{\Tr}{Tr}
\newcommand{\dd}{\mathrm d}
\newcommand{\CP}{c_{\mathrm P}}
\newcommand{\ket}[1]{\lvert#1\rangle}
\newcommand{\bra}[1]{\langle#1\rvert}
\newcommand{\norm}[1]{\lVert#1\rVert}
\begin{document}
\title{Critical properties of EPR steering}
\author{Yu-Xuan Zhang}
\affiliation{School of Physics, Nankai University, Tianjin 300071, People's Republic of China}
\author{Jing-Ling Chen}
\email{chenjl@nankai.edu.cn}
\affiliation{Theoretical Physics Division, Chern Institute of Mathematics, Nankai University, Tianjin 300071, People's Republic of China}
\date{October 30, 2026}
\begin{abstract}
Every PVM-critical entangled two-qubit state becomes steerable with two copies. We prove this for the full critical boundary and for two different critical inputs. The joint measurements can have arbitrarily little entanglement, while all conditional states on Bob's two systems remain separable. Criticality is exactly the point where this measurement-entanglement threshold vanishes. The number of measurements has a separate critical behavior: every fixed setting bound leaves an unsteerable neighborhood of the critical input. Moreover, for any such bound, there are critical states that remain unsteerable under every tensor power, although unrestricted measurements activate them with two copies. These results identify different roles for measurement entanglement, measurement number, and collective access in revealing steering. We also construct fixed-dimensional states whose steering activates on two copies while Bell locality persists through any prescribed finite copy range. In increasing dimension, the two-copy steering violation is unbounded, giving an affirmative solution to the amplification question of Hsieh, Liang, and Lee.
\end{abstract}
\maketitle
\clubpenalty=10000
\widowpenalty=10000
\brokenpenalty=10000

\section*{Introduction}

Every entangled two-qubit state at the boundary of unsteerability for projective measurements (PVMs) becomes steerable with two copies. We prove this throughout the boundary, including biased states, and also when the two critical inputs are different. The result identifies a class of mixed entangled states for which collective measurements reveal EPR steering at a uniform copy number.

Quantum steering originates in the Einstein--Podolsky--Rosen discussion of the completeness of quantum mechanics~\cite{EPR}. In his response, Schr\"odinger emphasized that measurements on one part of an entangled system can prepare different ensembles on the other part without changing its reduced state~\cite{Schrodinger,HJW}. Bell's theorem later established the conflict between quantum predictions and local hidden-variable theories~\cite{Bell}. Entanglement, steering, and Bell nonlocality nevertheless impose different conditions on a shared state. Werner constructed entangled states admitting a local model~\cite{Werner}, and Wiseman, Jones, and Doherty placed steering strictly between entanglement and Bell nonlocality~\cite{Wiseman,Jones}. This hierarchy remains strict when general measurements are allowed~\cite{QuintinoHierarchy}.

In a steering experiment, Alice seeks to convince Bob that their shared state is entangled, although Bob trusts only his own measurement devices. A local hidden-state (LHS) model explains all of Bob's conditional states by one fixed quantum ensemble and Alice's classical response probabilities. Steering occurs when no common ensemble can reproduce the experiment~\cite{Review,CavalcantiReview}. This requirement gives steering an operational role in one-sided device-independent quantum key distribution~\cite{Branciard} and subchannel discrimination~\cite{PianiWatrous}. It also makes the accessible measurements part of the resource question: the same state may admit an LHS model for one measurement class and fail to admit one for a larger class.

Steering inequalities provide experimentally accessible tests~\cite{CavalcantiCriteria,Saunders}, while semidefinite programming can detect and quantify steering for specified measurement families~\cite{Pusey,Skrzypczyk,CavalcantiReview}. The relation to measurement incompatibility expresses the common-ensemble requirement through a parent measurement~\cite{QuintinoJM,UolaJM,UolaCorrespondence,GuhneMeasurements}. For two qubits, geometric descriptions determine the projective boundary by comparing all required conditional states with the preparations available from a hidden-state ensemble~\cite{NguyenGeometry,Waves}. An exact boundary therefore contains information beyond a test for a particular set of measurements. It identifies states for which an LHS model exists but has no freedom to accommodate certain changes of the experiment.

Collective measurements expose this restriction. Hsieh, Liang, and Lee studied steering activation and the amplification of violations when copies are combined~\cite{Hsieh}. Quintino, Huber, and Brunner proved that every entangled two-qubit state becomes steerable at some finite copy number~\cite{QuintinoActivation}. That number may depend on the state. Our theorem shows that the entire entangled PVM boundary has the same activation number: two. The Werner-state case was given by Chitambar and Zhang~\cite{ChitambarZhang}. The general result follows because single-copy criticality constrains every possible hidden-state explanation of the two-copy experiment.

This raises a physical question about the activating measurement. If two independent inputs each admit an LHS model, how much entanglement must a joint measurement contain before their product becomes steerable? We show that the threshold is zero exactly at criticality. Below the boundary, a common model survives all sufficiently small departures from product measurements. At the boundary, arbitrarily weakly entangled measurements suffice. Yet their conditional states can all remain separable between Bob's two systems. For unbiased PVM-unsteerable inputs, every joint POVM has this remote separability property. Thus the entanglement needed in the measurement and the entanglement it prepares on the remote systems have different roles in activation.

The number of settings introduces another restriction. Experiments can improve their noise tolerance by using more settings~\cite{Saunders}, but every fixed setting bound leaves an unsteerable neighborhood of a critical input. The setting number must therefore grow without bound to detect single-copy steering arbitrarily close to the boundary. At criticality, all-PVM models also require infinitely many hidden states. Collective access does not generally remove the setting restriction: using symmetric extensions~\cite{Terhal}, we construct critical inputs for which any prescribed setting bound prevents steering at every copy number. Increasing the number of copies cannot always compensate for limited measurement choices.

The distinction between steering and other quantum resources remains relevant after activation. Steering with Bell-local states is already possible on one copy~\cite{Saunders,BowlesOneWay,QuintinoHierarchy}. We construct states that are unsteerable on one copy and steerable on two, while remaining Bell local throughout any prescribed finite copy range. Erasure preserves steering in the trusted direction and, together with noise, suppresses measurement incompatibility in the reverse direction~\cite{Sekatski}. The construction applies to the full-rank critical inputs covered by the known PVM--POVM equivalence results~\cite{ZhangUnital,Waves}. A separate construction uses high-dimensional isotropic states to give unbounded steering violations with two copies, providing the first affirmative solution to the two-copy amplification question of Ref.~\cite{Hsieh}.

Section~\ref{sec:criticality} introduces single-copy criticality, and Sec.~\ref{sec:activation} proves the activation theorem. Section~\ref{sec:phenomena} studies the entanglement thresholds of the measurements and inputs, together with remote separability. Section~\ref{sec:measurements} develops the consequences of finite measurement and hidden-state numbers. Section~\ref{sec:resources} treats Bell locality and the strength of the steering violation.

\section{Single-copy criticality}\label{sec:criticality}

Fix the steering direction $A\to B$. Alice's measurement effects $M_{a|x}$ prepare the unnormalized conditional states on Bob's side
\[
 \sigma_{a|x}=\Tr_A[(M_{a|x}\otimes I)\rho_{AB}], \qquad \sum_a\sigma_{a|x}=\rho_B.
\]
These conditional states admit a local hidden-state (LHS) model if they can be written as
\begin{equation}
 \sigma_{a|x}=\int p(a|x,\lambda)\tau_\lambda\,\dd\mu(\lambda),
 \label{lhs}
\end{equation}
where $\mu$ is a probability measure, $\tau_\lambda$ are normalized quantum states, and $p(a|x,\lambda)$ are conditional probabilities. An LHS model requires the hidden states and their weights to be independent of the measurement choice.

Now consider an entangled two-qubit state. Bob's reduced density matrix is necessarily full rank, so an invertible filter on Bob transforms it into $I/2$ and brings the shared state into the canonical form~\cite{NguyenGeometry}
\begin{equation}
 4\rho=I\otimes I+(\vec{a}\cdot\vec{\sigma})\otimes I+\sum_{ij}C_{ij}\sigma_i\otimes\sigma_j.
 \label{canonical}
\end{equation}
Here $\vec{\sigma}=(\sigma_x,\sigma_y,\sigma_z)$ is the Pauli vector, $\vec{a}$ is Alice's Bloch vector, and $C$ is the real correlation matrix. Since invertible filtering preserves LHS models, steerability is unchanged for each family of Alice's measurements~\cite{NguyenGeometry}. Alice's projective measurement along a unit vector $\vec{e}$ prepares the conditional states
\[
 \sigma_{\pm|\vec{e}}=\frac14[(1\pm\vec{a}\cdot\vec{e})I\pm(C^{\mathsf T}\vec{e})\cdot\vec{\sigma}]
\]
on Bob's side.

Write a hidden state as $\tau_{\vec{n}}=(I+\vec{n}\cdot\vec{\sigma})/2$, with $|\vec{n}|\le1$. For a measurement along $\vec{e}$, the difference between the response probabilities of the two outcomes,
$f_{\vec{e}}=p(+|\vec{e},\vec{n})-p(-|\vec{e},\vec{n})$, must satisfy $|f_{\vec{e}}|\le1$ and
\[
 \int f_{\vec{e}}(\vec{n})(1,\vec{n})\,\dd\mu(\vec{n})=(\vec{a}\cdot\vec{e},C^{\mathsf T}\vec{e}).
\]
Thus the same hidden-state ensemble must reproduce the conditional states required for every direction $\vec{e}$. This can be expressed as an inclusion relation between two convex sets~\cite{NguyenGeometry}. For a linear test $(s,\vec{y})$, maximizing over all allowed responses gives $K_\mu(s,\vec{y})$, while maximizing over all measurement directions gives $N(s,\vec{y})$, where
\[
 \begin{aligned}
 K_\mu(s,\vec{y})&=\int|s+\vec{y}\cdot\vec{n}|\,\dd\mu(\vec{n}),\\[2pt]
 N(s,\vec{y})&=\max_{|\vec{e}|=1}\vec{e}\cdot(\vec{a}s+C\vec{y})=|\vec{a}s+C\vec{y}|.
 \end{aligned}
\]
The ensemble reproduces all measurement outcomes if and only if $K_\mu\ge N$. Following Ref.~\cite{Waves}, let $\CP$ denote the minimum total weight of a positive measure satisfying
\begin{equation}
 \int\vec{n}\,\dd\mu=0,\qquad K_\mu(s,\vec{y})\ge N(s,\vec{y}) \quad\text{for all }(s,\vec{y}).
 \label{mass}
\end{equation}
If its total weight is below one, the remaining weight can be placed at the maximally mixed state with equal response probabilities for the two outcomes to construct an LHS model. Thus a state is unsteerable for all PVMs if and only if $\CP\le1$.

When an entangled state satisfies $\CP=1$, it is at the threshold of unsteerability, and we call it a PVM-critical state. Along the white-noise path $\rho_t=t\rho+(1-t)I_4/4$, homogeneity gives $\CP(\rho_t)=t\CP(\rho)$. Thus this definition also identifies the boundary of PVM unsteerability along the white-noise direction.

We now give two properties used below: 1. For an entangled state, $C$ is invertible. Otherwise, Bob's coordinates can be rotated so that the $y$ column of $C$ vanishes. The canonical state is then invariant under partial transposition on Bob and hence separable by the two-qubit PPT criterion~\cite{Peres,Horodecki}, contradicting entanglement. 2. For a critical state, Eq.~(S181) of Ref.~\cite{Waves} gives
\begin{equation}
 \alpha:=\max_{|\vec{n}|\le1}|\vec{a}+C\vec{n}|<\CP=1.
 \label{gap}
\end{equation}
In particular, all conditional states prepared by local projective measurements are strictly positive, and their smallest eigenvalues have a common positive lower bound independent of the measurement direction.

\section{Two-copy activation at criticality}\label{sec:activation}

\begin{theorem}\label{thm:activation}
Let $\rho_1$ and $\rho_2$ be PVM-critical entangled two-qubit states from Alice to Bob. Then $\rho_1\otimes\rho_2$ is PVM-steerable in the same direction.
\end{theorem}

Taking $\rho_1=\rho_2$ gives two-copy activation. Bob's invertible filters preserve steerability, so we use the canonical form~\eqref{canonical}. We first apply the single-copy constraint $K_\mu\ge N$ to the critical inputs.

By homogeneity, tests can be normalized to $s^2+|\vec{y}|^2=1$. Nonnegative test weights $\lambda$ satisfying
\begin{equation}
 \int|s+\vec{y}\cdot\vec{n}|\,\dd\lambda(s,\vec{y})
 \le1+\vec{b}\cdot\vec{n},\qquad |\vec{n}|\le1,
 \label{dualbound}
\end{equation}
for some vector $\vec{b}$ give the lower bound
\[
 \int N\,\dd\lambda\le\int K_\mu\,\dd\lambda\le\int\dd\mu.
\]
The last inequality uses the zero mean of $\mu$. Convex duality gives an attained maximum $\int N\,\dd\lambda=\CP$. At criticality, both ends are one for every LHS model. Consequently,
\begin{equation}
 K_\mu(s,\vec{y})=N(s,\vec{y})
 \label{saturation}
\end{equation}
on the support of an optimal $\lambda$, and Eq.~\eqref{dualbound} is an equality on the support of every LHS ensemble.

The supported tests have $|s|<|\vec{y}|$ and $N>0$. Otherwise, $|s|\ge|\vec{y}|$ would give $N\le\alpha|s|<K_\mu$ by Eq.~\eqref{gap}; $N=K_\mu=0$ would imply $s=0$ and $C\vec{y}=0$, contrary to invertibility and normalization. Compactness then bounds $|\vec{y}|^2-s^2$ and $N$ away from zero.

For the direction $\vec{e}=(\vec{a}s+C\vec{y})/N(s,\vec{y})$, the moment condition in Sec.~\ref{sec:criticality} gives
\begin{equation}
 \begin{aligned}
 \int f_{\vec{e}}(\vec{n})(s+\vec{y}\cdot\vec{n})\,\dd\mu&=N(s,\vec{y})\\
 &=\int|s+\vec{y}\cdot\vec{n}|\,\dd\mu.
 \end{aligned}
 \label{fixedresponse}
\end{equation}
Since $|f_{\vec{e}}|\le1$, the response is fixed to $\operatorname{sgn}(s+\vec{y}\cdot\vec{n})$ wherever the test is nonzero. The hidden states must also be pure. Equality in Eq.~\eqref{dualbound} at an interior point would, by convexity, force equality throughout the Bloch ball. But $|s|<|\vec{y}|$ makes the spherical average of the left side strictly exceed its center value, while the right side has equal average and center value. This is impossible.

Now suppose all product PVMs and the joint measurements below share an LHS model. Each hidden marginal must be pure by the single-copy result. Hence every joint hidden state is a product $\tau_{\vec{n}_1}\otimes\tau_{\vec{n}_2}$; write its distribution as $\nu$. For a supported test of each input, set
\[
 H_i=s_iI+\vec{y}_i\cdot\vec{\sigma},\qquad
 \vec{e}_i=\frac{\vec{a}_i s_i+C_i\vec{y}_i}{N_i(s_i,\vec{y}_i)}.
\]
Here $N_i$ is $N$ for $\rho_i$. The two local outcomes are fixed by their test signs, and their parity satisfies
\[
 \int\prod_{i=1}^2|s_i+\vec{y}_i\cdot\vec{n}_i|\,\dd\nu
 =N_1(s_1,\vec{y}_1)N_2(s_2,\vec{y}_2).
\]

Rotate the product basis by
\begin{equation}
 U_\theta=\exp[-i\theta(\sigma_x\otimes\sigma_x)],
 \label{gate}
\end{equation}
with $\sigma_x$ expressed in these local bases. The rotation preserves the even and odd subspaces, so the parity bound remains saturated. Except at zero test values, an odd outcome can therefore respond only to opposite signs. Consider
\begin{equation}
 \ket{\psi_\theta}=\cos\theta\ket{+_{\vec{e}_1}-_{\vec{e}_2}}
 -i\sin\theta\ket{-_{\vec{e}_1}+_{\vec{e}_2}},
 \label{rotated}
\end{equation}
and denote its conditional state by $\sigma_\theta$ and its response change by $\Delta=p_\theta-p_0$.

The new coherence changes the projector to first order in $\theta$. This change survives on Bob: the canonical map takes Alice's Pauli coefficients $(x_0,\vec{x})$ to $(x_0+\vec{a}_i\cdot\vec{x},C_i^{\mathsf T}\vec{x})/2$, and is invertible because $C_i$ is invertible. Its product therefore gives, for a uniform $c>0$,
\begin{equation}
 \int|\Delta|\,\dd\nu\ge\norm{\sigma_\theta-\sigma_0}_1\ge c\theta
 \label{firstorder}
\end{equation}
at sufficiently small positive $\theta$.

The model cannot supply this first-order change freely. At signs $(+,-)$, the original response is one and can only decrease; at $(-,+)$, it is zero and can only increase. At equal signs it stays zero. Thus, with
\[
 G=H_1\otimes I-I\otimes H_2,\qquad
 v=\sum_{i=1}^2|s_i+\vec{y}_i\cdot\vec{n}_i|,
\]
both allowed changes contribute with the same sign:
\begin{equation}
 \Tr[G(\sigma_\theta-\sigma_0)]=-\int|\Delta|v\,\dd\nu.
 \label{responsecost}
\end{equation}
The quantum value, however, changes only to second order. Indeed,
\[
 \Tr_{B_i}[(I\otimes H_i)\rho_i]
 =\frac12[s_iI+N_i(s_i,\vec{y}_i)\vec{e}_i\cdot\vec{\sigma}]
\]
is diagonal in the chosen basis. Each term obtained by transferring $G$ to Alice contains such a diagonal factor, so its matrix element between $\ket{+-}$ and $\ket{-+}$ vanishes. Only the $\sin^2\theta$ terms remain, giving a uniform $D>0$ with
\begin{equation}
 \int|\Delta|v\,\dd\nu\le D\theta^2.
 \label{cost}
\end{equation}

The two bounds require the response change to concentrate where both tests are close to zero. Criticality limits the weight available there:
\begin{equation}
 \lambda\{(s,\vec{y}):|s+\vec{y}\cdot\vec{n}|\le\varepsilon\}
 \le B\varepsilon,
 \label{strip}
\end{equation}
uniformly over the allowed hidden states. To see this, take a unit $\vec{u}\perp\vec{n}$ and add Eq.~\eqref{dualbound} at $\sqrt{1-t^2}\vec{n}\pm t\vec{u}$, subtracting twice its equality at $\vec{n}$. The radial displacement is of order $t^2$. The identity $|x+y|+|x-y|=2\max\{|x|,|y|\}$ then gives
\[
 \int\bigl(t|\vec{y}\cdot\vec{u}|-|s+\vec{y}\cdot\vec{n}|\bigr)_+\,\dd\lambda\le At^2,
\]
where $(x)_+=\max\{x,0\}$. Near a zero test value, the lower bound on $|\vec{y}|^2-s^2$ keeps the tangential component of $\vec{y}$ nonzero. Taking two orthogonal tangent directions and $t$ proportional to $\varepsilon$ proves Eq.~\eqref{strip}. Exact zeros consequently have no weight in the test average, which also justifies using Eqs.~\eqref{responsecost} and \eqref{cost} in that average.

Normalize the two test measures and average independently. For each hidden-state pair, Eq.~\eqref{strip} bounds the weight of $v\le\varepsilon$ by $B_1B_2\varepsilon^2$. On $v>\varepsilon$, Eq.~\eqref{cost} bounds the response change by $D\theta^2/\varepsilon$. Combining with Eq.~\eqref{firstorder} gives
\begin{equation}
 c\theta\le B_1B_2\varepsilon^2+\frac{D\theta^2}{\varepsilon},
 \label{contradiction}
\end{equation}
where the constants include the normalization factors. Set $\varepsilon=\theta^{2/3}$. The right side is of order $\theta^{4/3}$ and cannot reproduce the first-order change on the left. Hence every sufficiently small positive $\theta$ excludes an LHS model. By compactness, a finite subfamily already witnesses steering.

\section{Critical phenomena and entanglement thresholds}\label{sec:phenomena}

A product measurement on two PVM-unsteerable inputs is described by the product of their LHS models. Theorem~\ref{thm:activation} shows that, for critical inputs, an arbitrarily small entangling rotation destroys this description. We first establish that criticality is exactly the condition for this zero threshold.

For a rank-one two-qubit PVM $M$, let $E(M)$ be the largest concurrence of its basis vectors. Thus $E(M)=0$ exactly for a product basis.

\begin{corollary}\label{cor:equivalence}
An entangled PVM-unsteerable two-qubit state is critical if and only if its two copies are steerable by joint PVMs with arbitrarily small $E(M)$.
\end{corollary}

At criticality, the bases in Theorem~\ref{thm:activation} have concurrence $\sin2\theta$, which can be arbitrarily small. To prove the converse, take $\CP<1$. Complete an optimal hidden-state ensemble with weight $1-\CP$ at $I/2$, assigning equal probabilities to both outcomes. The product ensemble simulates all orthogonal product bases on two copies. This includes bases implemented by choosing the second local measurement according to the first outcome: every orthogonal product basis of two qubits admits such a local procedure. Indeed, after one basis vector is made $\ket{00}$, two of the remaining vectors must share a $\ket1$ factor on the same side; the other two vectors form the complementary branch.

The product model has a maximally mixed component of weight $(1-\CP)^2$, with four conditional states $I_4/16$. This component gives the responses freedom to change. Let $\tau_j$ be the four pure qubit states at the vertices of a regular tetrahedron. For each outcome,
\begin{equation}
 \sigma_a^0=\frac1{16}\sum_{j,k}p(a|j,k)\tau_j\otimes\tau_k,
 \qquad p(a|j,k)=\frac14.
 \label{tetrahedron}
\end{equation}
The sixteen products form a basis of Hermitian operators. Any sufficiently small change of the conditional states can therefore be reproduced by small changes of these positive responses. The fixed Bob marginal keeps the responses normalized. All other parts of the model can be left unchanged.

Consequently, one common ensemble simulates a neighborhood of every product basis. Compactness of the product bases makes this neighborhood uniform: some $\varepsilon>0$ ensures an LHS model for all joint PVMs with $E(M)<\varepsilon$. This proves the converse.

Along the white-noise path $\rho_t$ of Sec.~\ref{sec:criticality}, this threshold approaches zero as $t\uparrow1$. To see this, choose any sufficiently weak joint rotation that witnesses steering at $t=1$. Its strict violation survives a small amount of input noise. Since the rotation can be arbitrarily weak, the positive thresholds at $t<1$ cannot remain bounded away from zero. Product measurements themselves remain unsteerable at the endpoint.

The entanglement in the activating measurement need not create entanglement between Bob's systems. Equation~\eqref{gap} gives a common positive lower eigenvalue bound for all local conditional states. At a product measurement, the joint conditional state and its partial transpose are therefore strictly positive. Both remain positive under a sufficiently small joint rotation. By the two-qubit PPT criterion, all conditional states in the activating construction can be separable across $B_1:B_2$~\cite{Peres,Horodecki}. Bob's local filters preserve this separability.

For unbiased canonical inputs, the remote separability statement extends to every joint POVM. In this case~\cite{Waves},
\begin{equation}
 \CP(\vec{0},C)=2\int_{S^2}|C\vec{u}|\,\dd\omega(\vec{u})\ge\norm{C}_F,
 \label{unbiased}
\end{equation}
where $\omega$ is normalized surface measure and $\norm{C}_F^2$ is the sum of squared singular values. To obtain the inequality, write those squares as $S^2p_i$, with $\sum_i p_i=1$. Then $\sqrt{\sum_i p_i u_i^2}\ge\sum_i p_i|u_i|$ and $\int|u_i|\,\dd\omega=1/2$. PVM unsteerability thus implies $\norm{C}_F^2\le1$.

Let $\mathcal E_i$ be the unital qubit channels with Choi states $\rho_i$. Their Bloch matrices have the same singular values as $C_i$. Under the bound just obtained, $\mathcal E_1\otimes\mathcal E_2$ maps every two-qubit state to a separable state~\cite{Filippov}. An Alice effect $M$ prepares
\[
 \sigma_M=\frac14(\mathcal E_1\otimes\mathcal E_2)(M^{\mathsf T}),
\]
so every joint POVM outcome is separable. Critical inputs nevertheless give steering by Theorem~\ref{thm:activation}. Each conditional state separately admits a separable decomposition, but these decompositions cannot be combined into one ensemble with responses for all measurements. This is the obstruction that the steering experiment detects.

We next consider the entanglement of the input. The critical boundary has a uniform two-copy activation number. Close to the separable boundary, however, entanglement can remain undetected by collective steering experiments over any prescribed finite copy range.

\begin{proposition}\label{prop:copybound}
For every finite $n_0$, some entangled two-qubit state is unsteerable for all POVMs on each of its first $n_0$ tensor powers.
\end{proposition}

Choose an entangled state $\rho_{\mathrm u}$ that is unsteerable for all POVMs~\cite{ZhangUnital,Waves}, and an interior separable state $\sigma_0$. Let $\sigma_1$ be the last separable point on the line from $\sigma_0$ to $\rho_{\mathrm u}$, so that
\[
 \rho_{\mathrm u}=(1+\beta)\sigma_1-\beta\sigma_0,\qquad \beta>0.
\]
For $0<\epsilon\le\beta$, the states $\varrho_\epsilon=(1+\epsilon)\sigma_1-\epsilon\sigma_0$ are physical and entangled. We prove unsteerability at $n_0$ for small $\epsilon$; discarding copies gives the lower powers.

Mixtures with one factor $\rho_{\mathrm u}$ and all other factors separable are unsteerable for joint POVMs: the separable ancillary state induces a POVM on the unsteerable factor. Let $S_j$ be the permutation average of $n_0-j$ factors $\sigma_1$ and $j$ factors $\sigma_0$. Replacing one $\sigma_1$ by $\rho_{\mathrm u}$ gives an unsteerable state
\[
 T_j=(1+\beta)S_j-\beta S_{j+1}.
\]
Expand $\varrho_\epsilon^{\otimes n_0}=\sum_jc_jS_j$, where $c_j=\binom{n_0}{j}(1+\epsilon)^{n_0-j}(-\epsilon)^j$. Each negative odd term can be paired with the preceding positive even term. For even $j<n_0$, put $r_j=-c_{j+1}/\beta>0$; then
\begin{equation}
 c_jS_j+c_{j+1}S_{j+1}
 =[c_j-(1+\beta)r_j]S_j+r_jT_j.
 \label{pairing}
\end{equation}
The ratio $r_j/c_j$ tends to zero with $\epsilon$, so all weights are nonnegative for sufficiently small $\epsilon$. Any unpaired final term is positive, and the weights sum to one. This gives the required LHS decomposition.

Every fixed entangled two-qubit state still activates at some finite copy number~\cite{QuintinoActivation}. Proposition~\ref{prop:copybound} shows that this number cannot be chosen uniformly over all entangled inputs. Fixing the input first and fixing the available copy number first therefore lead to different conclusions. The construction places the latter inputs arbitrarily close to separability, while critical inputs always activate with two copies.

\section{Criticality and the number of measurement settings}\label{sec:measurements}

The projective boundary concerns all PVMs, whereas an experiment uses finitely many. We examine the consequences for the hidden-state ensemble and for the number of settings needed to reveal steering.

\begin{corollary}\label{cor:hiddenstates}
An entangled PVM-unsteerable two-qubit state is critical if and only if every LHS model for all PVMs contains infinitely many hidden states.
\end{corollary}

At criticality, suppose an LHS ensemble is finite. Equation~\eqref{strip} then guarantees a saturated test that is nonzero at every hidden state. Its signs remain unchanged under a small perturbation, so $K_\mu$ is locally linear. Choose $\vec{u}$ such that $C\vec{u}\ne0$ and $C\vec{u}\perp\vec{a}s+C\vec{y}$. At the saturated test,
\begin{equation}
 N(s,\vec{y}+r\vec{u})
 =\sqrt{N(s,\vec{y})^2+r^2|C\vec{u}|^2}.
 \label{curvature}
\end{equation}
A linear function equal to this curve at $r=0$ cannot stay above it on both sides. Thus a finite ensemble cannot satisfy $K_\mu\ge N$.

For $\CP<1$, a finite model can be constructed using the unused weight~\cite{BowlesRandom,HirschModels}. Add a small uniform spherical measure to an optimal measure, keeping the total weight below one. Every normalized constraint then has a common positive margin. Partition the Bloch ball into sufficiently small cells and concentrate each cell's weight at its mean vector. This preserves the total weight and zero mean while changing $K_\mu$ by less than the margin. The resulting finite measure remains feasible. Completing it at $I/2$ gives a finite LHS model for all PVMs, proving the converse.

Along the path $\rho_t$ towards a critical state, such finite models exist for every $t<1$, but their size diverges as $t\uparrow1$. Otherwise, a convergent subsequence of models with bounded size would give a finite model at the critical endpoint. The infinite ensemble is needed to describe all PVM directions simultaneously. Any fixed finite set of unsteerable PVMs has a finite LHS model, obtained by refining the responses into deterministic outcome assignments.

To connect this ensemble requirement to finite experiments, consider a fixed number of measurement settings.

\begin{proposition}\label{prop:finitesettings}
For every finite $m$, each PVM-critical entangled two-qubit state has a neighborhood unsteerable under all PVM families with at most $m$ settings.
\end{proposition}

For a finite set of directions $F$, the moment constraints require only
\[
 N_F(s,\vec{y})=\max_{\vec{e}\in F}|\vec{e}\cdot(\vec{a}s+C\vec{y})|.
\]
Write $\CP(\rho;F)$ for the corresponding minimum weight in Eq.~\eqref{mass}. We will show that these finitely many directions leave a strict margin:
\begin{equation}
 \CP(\rho;F)<1.
 \label{finitegap}
\end{equation}

The curvature argument above also prevents an optimal all-PVM test measure $\lambda$ from having an atom. A positive weight at one test would, by Eq.~\eqref{strip}, keep that test uniformly away from zero on every allowed hidden state. The fixed signs again make $K_\mu$ locally linear at a point where it touches the curved function $N$.

For finite $F$, however, an optimal test measure can be replaced by finitely many tests. Group tests according to the signed direction attaining $N_F$, and replace each group by its weighted mean. Convexity preserves the dual inequality~\eqref{dualbound}, while linearity within each group preserves the objective. Homogeneity renormalizes each nonzero mean. If $\CP(\rho;F)=1$, the resulting finite measure would also be optimal for all PVMs, contradicting the absence of atoms. This proves Eq.~\eqref{finitegap}.

As in the finite-ensemble construction, a small spherical component turns the remaining weight into a positive margin on all normalized tests. This margin survives small changes of the input and the directions. The space of $m$ directions is compact, so one input neighborhood works for every family with at most $m$ settings. Bob's canonical filter varies continuously near the full-rank marginal, giving the same conclusion for the original states.

Consequently, the minimum number of settings diverges whenever PVM-steerable states approach a critical entangled state. Each steerable input has a finite witness, but a bound fixed in advance fails sufficiently close to criticality. For example, if the radial continuation $\rho_r=r\rho+(1-r)I_4/4$ is physical for $r>1$, then $\CP(\rho_r)=r$ and these states are steerable. For every fixed $m$, their steering becomes undetectable with $m$ settings as $r\downarrow1$. The endpoint itself has an LHS model for all PVMs. This distinguishes choosing the measurements after the input is specified from fixing an experimental setting bound before approaching the boundary.

The same restriction appears in the two-copy experiment when the critical input is held fixed.

\begin{corollary}\label{cor:settings}
For two copies of a fixed PVM-critical entangled two-qubit state, the number of joint PVM settings required for steering diverges as $\max_M E(M)\to0$.
\end{corollary}

Each finite family of product bases uses finitely many local directions. Equation~\eqref{finitegap} leaves a nonzero maximally mixed component in the LHS models for these directions. The product component supplies the positive responses in Eq.~\eqref{tetrahedron}, and hence an unsteerable neighborhood of that product family. If steerable families with bounded setting number approached product bases, a convergent subsequence would enter such a neighborhood. This contradiction proves the corollary.

Each nonzero rotation in Theorem~\ref{thm:activation} admits a finite witness, but the setting number must grow as the rotation vanishes. Thus the zero entanglement threshold comes with an unbounded requirement on measurement choices.

One might then expect additional copies to compensate for limited settings. Symmetric extensions show that this need not happen.

\begin{proposition}\label{prop:settingcopies}
For every finite $m$, some PVM-critical entangled two-qubit state is unsteerable with at most $m$ POVM settings on every tensor power.
\end{proposition}

An extension to $m$ interchangeable Alice systems, each sharing the original state with Bob, provides an LHS model for any $m$ measurements~\cite{Terhal}. Measure one POVM on each extension system; the joint outcomes and conditional Bob states supply the model. Tensor powers preserve this property: grouping the same extension index across copies gives an $m$-fold extension of the tensor power. Thus it remains to construct a critical entangled state with such an extension.

For $m\ge2$, take
\begin{equation}
 \begin{aligned}
 4\zeta_t={}&I\otimes I-t(\sigma_x\otimes\sigma_x+\sigma_y\otimes\sigma_y)\\
 &-c(t)\sigma_z\otimes\sigma_z,
 \end{aligned}
 \label{criticalfamily}
\end{equation}
where $0<t<1/2$ and $c(t)\in(t,1)$ is determined by
\begin{equation}
 2\int_0^1\sqrt{t^2+[c(t)^2-t^2]u^2}\,\dd u=1.
 \label{criticalroot}
\end{equation}
The integral is strictly increasing in $c$ and crosses one between $t$ and $1$. Equation~\eqref{unbiased} gives $\CP=1$. The eigenvalues $(1\pm2t+c)/4$ and $(1-c)/4$ twice are positive. The strict triangle inequality in Eq.~\eqref{unbiased} gives $1<2t+c$, so the state is entangled. Direct integration yields
\[
 1-c(t)=t^2\log(1/t)+O(t^2)\qquad(t\downarrow0).
\]

To obtain the extension, construct a channel with $m$ symmetric outputs whose single-output Bloch matrix is $\operatorname{diag}(t,t,c(t))$. Let $\ket{D_m^j}$ denote the normalized symmetric Dicke state with $j$ excitations. The isometry
\begin{equation}
 \begin{aligned}
 V\ket0&=\sqrt{1-q}\,\ket{0^m}\ket0_E+\sqrt q\,\ket{D_m^{m-1}}\ket1_E,\\[2pt]
 V\ket1&=\sqrt{1-q}\,\ket{1^m}\ket1_E+\sqrt q\,\ket{D_m^1}\ket0_E
 \end{aligned}
 \label{broadcast}
\end{equation}
has symmetric outputs with marginal Bloch matrix $\operatorname{diag}(t,t,c_m(t))$, where
\[
 q=\frac{1-\sqrt{1-mt^2}}2,\qquad
 c_m(t)=1-\frac{2(m-1)}m q.
\]
The off-diagonal terms give $t=2\sqrt{q(1-q)/m}$, and the excitation probabilities give $c_m(t)$.

A second channel mixes complete dephasings along $x,y,z$ with weights $t,t,1-2t$. It measures and prepares states, so it can produce arbitrarily many identical outputs. Its marginal matrix is $\operatorname{diag}(t,t,1-2t)$. For fixed $m$ and sufficiently small $t$,
\[
 1-2t<c(t)<c_m(t),
\]
because $1-c_m(t)=O(t^2)$. A mixture of the two channels therefore has the required marginal. Applying it to Alice's half of a singlet gives the $m$-fold extension of $\zeta_t$. The case $m=1$ is immediate.

Along $t\downarrow0$, the family stays critical and approaches a separable state. For any prescribed setting bound, one member remains unsteerable under that bound for every copy number, although its two copies are steerable with enough settings. The uniform activation number therefore leaves the measurement requirement unbounded.

\section{Further resource consequences of criticality}\label{sec:resources}

\subsection{Activated Bell-local states}\label{sec:erasure}

Bell nonlocality requires both parties to exclude a local explanation using only their observed statistics. Steering uses the additional knowledge that the trusted party's conditional states are quantum states. This difference permits steering activation while Bell locality persists. The following construction combines the two-copy theorem with the directionality of steering~\cite{BowlesOneWay,QuintinoHierarchy}.

\begin{corollary}\label{cor:bell}
For every finite $n_0\ge2$, there is a $2\times3$ state that is unsteerable for all POVMs on one copy, steerable on two copies, and Bell local for all joint POVMs through $n_0$ copies.
\end{corollary}

We use erasure on Bob. For a full-rank state $\rho_{AB}$ and a flag $\ket e$ orthogonal to Bob's input space, define
\begin{equation}
 \Omega_\eta=\eta\rho+(1-\eta)\rho_A\otimes\ket e\bra e,
 \qquad 0<\eta\le1.
 \label{erasure}
\end{equation}
Bob can select the surviving sector when his devices are trusted. In the reverse direction, the same loss weakens the measurements used to steer Alice.

\begin{proposition}\label{prop:erasure}
For the full-rank input in Eq.~\eqref{erasure}, erasure preserves $A\to B$ steerability on every tensor power for each Alice measurement family. For any finite copy bound, sufficiently small positive $\eta$ makes the reverse direction unsteerable for all POVMs throughout that range.
\end{proposition}

Erasure preserves every LHS model. Conversely, Bob can project onto the sector in which all copies survive, recovering the original assemblage with probability $\eta^n$. This trusted filter also preserves LHS models, giving forward equivalence for the same Alice measurements.

For the reverse direction, full rank gives $0<\ell<1$ such that $\rho\ge\ell\rho_A\otimes I_{d_B}/d_B$. Each $k$-copy block can consequently be written as
\[
 \rho^{\otimes k}=(\mathrm{id}\otimes\mathcal D_{1-\ell^k})(\widetilde\rho_k),
\]
where $\widetilde\rho_k$ is a state and $\mathcal D_p(X)=pX+(1-p)\Tr(X)I/d_B^k$. Depolarization followed by erasure makes all effective POVMs jointly measurable when the survival probability satisfies $q\le(1-p)^{d-1}$~\cite{Sekatski}. A parent measurement then supplies the hidden outcomes and conditional Alice states of a reverse LHS model. For a globally erased $k$-copy block, it is sufficient to take
\[
 q_k=\ell^{k(d_B^k-1)}>0.
\]

Independent erasure has several surviving sectors. For a fixed $n_0$, there are finitely many, and each can be combined with part of the all-erased sector to obtain the globally erased block just described. Let $R_S$ contain $\rho$ at the positions $S\subseteq\{1,\ldots,n_0\}$ and $\rho_A\otimes\ket e\bra e$ at the other positions. For $S\ne\varnothing$,
\[
 T_S=q_{|S|}R_S+(1-q_{|S|})R_\varnothing
\]
is reverse unsteerable. The weight of $R_S$ in $\Omega_\eta^{\otimes n_0}$ is $p_S=\eta^{|S|}(1-\eta)^{n_0-|S|}$. With $r_S=p_S/q_{|S|}$,
\begin{equation}
 \Omega_\eta^{\otimes n_0}
 =\sum_{S\ne\varnothing}r_ST_S
 +\left(1-\sum_{S\ne\varnothing}r_S\right)R_\varnothing.
 \label{erasuredecomposition}
\end{equation}
All $r_S$ tend to zero as $\eta\downarrow0$. For sufficiently small positive $\eta$, this is a convex mixture of reverse-unsteerable states. Discarding copies gives the same result for every $n\le n_0$.

To prove Corollary~\ref{cor:bell}, choose a full-rank PVM-critical entangled two-qubit input that is also unsteerable for all POVMs. The PVM and POVM thresholds coincide for all Bell-diagonal states~\cite{ZhangUnital}; Theorem~S24 of Ref.~\cite{Waves} covers this class and further biased states. The family~\eqref{criticalfamily} provides full-rank examples. Its forward two-copy steering survives erasure by Theorem~\ref{thm:activation} and Proposition~\ref{prop:erasure}. The reverse LHS model implies Bell locality: Alice's Born probabilities become her local response probabilities for any POVM. Bell locality therefore holds through $n_0$ copies. The one-copy state remains forward unsteerable under Bob's erasure and is reverse unsteerable by construction, completing the proof.

The construction applies to any input with the stated properties. Local erasure on Bob's systems also preserves the separable conditional states of Sec.~\ref{sec:phenomena}. Thus two-copy steering, remote separability, and Bell locality coexist in fixed dimension throughout a prescribed finite copy range.

\subsection{Unbounded violations from two copies}\label{sec:amplification}

Activation asks whether combining copies produces a violation. Its strength is a further resource question. Hsieh, Liang, and Lee obtained unbounded amplification with three copies and asked whether two suffice~\cite{Hsieh}. Isotropic states at their single-copy PVM boundary answer this question affirmatively.

\begin{corollary}\label{cor:amplification}
Two copies of PVM-unsteerable states can violate steering inequalities by an arbitrarily large factor.
\end{corollary}

Take the boundary isotropic state~\cite{Wiseman,Jones}
\[
 \omega_d=p_d\Phi_d+\frac{1-p_d}{d^2}I_{d^2},\qquad
 p_d=\frac{H_d-1}{d-1},
\]
where $\Phi_d$ is the maximally entangled state projector and $H_d=\sum_{j=1}^d1/j$. Its overlap with $\Phi_d$ is $f_d=[(d+1)H_d-d]/d^2$. Two copies have local dimension $D=d^2$ and overlap $f_d^2$ with $\Phi_D$.

The overlap can be tested by averaging over orthonormal bases. For $P_{a|U}=U\ket a\bra aU^\dagger$, define
\[
 S_D(\sigma)=\int\dd U\sum_a\Tr(P_{a|U}\sigma_{a|U}),
\]
with normalized Haar measure. A pure hidden state obtains its largest score by choosing the basis vector with the largest squared overlap. The average of that maximum is $H_D/D$~\cite{Wiseman,Jones}, which by convexity also bounds mixed hidden states. Hence every LHS model satisfies $S_D\le H_D/D$.

Alice's measurements $P_{a|U}^{\mathsf T}$ give the quantum score $\Tr(W_D\rho)$, with
\begin{equation}
 W_D=\int\dd U\sum_aP_{a|U}^{\mathsf T}\otimes P_{a|U}
 =\frac{I_{D^2}+D\Phi_D}{D+1}.
 \label{haaroperator}
\end{equation}
Thus the test depends only on the maximally entangled overlap and applies directly to $\omega_d^{\otimes2}$. The ratio of the quantum score to the LHS bound is $D(1+Df)/[(D+1)H_D]$. Substituting $D=d^2$ and $f=f_d^2$ gives
\begin{equation}
 \begin{aligned}
 V_d&=\frac{[(d+1)H_d-d]^2+d^2}{(d^2+1)H_{d^2}}\\
 &=\tfrac12\log d+O(1)\longrightarrow\infty.
 \end{aligned}
 \label{amplification}
\end{equation}
The two-copy quantum score scales as $\log^2d/d^2$, while the LHS bound scales as $\log d/d^2$. Their ratio is therefore unbounded. Finite settings suffice for every desired violation: at fixed dimension, continuity gives a finite weighted set of bases approximating the quantum score and the LHS bound uniformly. Choose $d$ with $V_d$ above the desired ratio, then choose a sufficiently accurate finite approximation. The copy number stays equal to two as the violation grows.

\section*{Discussion and conclusions}

The entangled two-qubit PVM boundary has a uniform activation property: two critical inputs become steerable when measured jointly, whether or not the inputs are identical. This identifies a complete class of mixed states for which single-copy unsteerability ceases at the first possible increase in copy number. The distinction from eventual activation is substantial. Every fixed entangled two-qubit input can activate at some finite copy number, but no finite number works for all entangled inputs. Near separability, unsteerability can persist throughout any range chosen in advance. At the steering boundary, two copies always suffice.

The measurement-entanglement threshold describes how the two-copy steering becomes accessible. Below criticality, the LHS model can absorb small changes of product measurements. At criticality, arbitrarily weak entangling rotations are sufficient. This identifies the boundary through the response to a change of the experiment. It also separates the role of the joint measurement from the state it prepares: every conditional state in the activating experiment can remain separable between Bob's systems. For unbiased PVM-unsteerable inputs, this holds for every joint POVM. The resource is witnessed by the incompatibility of all these preparations with one common ensemble, even though each preparation separately admits a separable description.

The description of the entire PVM class has a corresponding consequence for finite resources. Critical inputs require infinitely many hidden states, whereas subcritical inputs admit finite models. Finite measurement families retain a strict margin at criticality, so they cannot resolve single-copy steering arbitrarily close to the boundary. In the two-copy experiment, the number of settings also grows as the activating measurements approach product measurements. These statements specify how a boundary defined by an unrestricted measurement class is approached by finite experiments. They also explain why a finite witness for each input does not supply one fixed setting bound for the whole class.

Additional copies do not generally remove the setting restriction. A symmetric extension can support an LHS model for a fixed number of measurements on every tensor power. We constructed critical states with extensions of any prescribed order, although their unrestricted two-copy experiments are steerable. Thus copies and settings provide different forms of access to quantum correlations. The order of specifying an input and imposing an experimental bound matters for both resources, with different behavior near the steering and separable boundaries.

The two further consequences concern the hierarchy and strength of nonclassical correlations. Erasure retains the forward steering experiment while giving the reverse direction an LHS model over a prescribed finite copy range. This yields activated Bell-local states in fixed dimension and preserves remote separability. In higher dimensions, isotropic states show that two-copy activation can also have unbounded violation strength. The affirmative solution to the two-copy amplification question uses a test determined only by the maximally entangled overlap, making the difference between its quantum and LHS scaling explicit.

Several questions remain. The critical construction guarantees finite witnesses, and finding economical measurement families would make its experimental requirements more explicit. Extending the activation theorem to higher dimensions requires an alternative to the pure-hidden-marginal constraint used for qubits. The erasure construction permits one state for each prescribed finite range; whether an entangled state can remain Bell local under all tensor powers remains a separate question. These directions concern how the boundary properties established here extend when dimension, collective access, and experimental restrictions are changed.

\begin{acknowledgments}
This work is supported by the Quantum Science and Technology--National Science and Technology Major Project (Grant No. 2024ZD0301000), and the National Natural Science Foundation of China (Grant No. 12275136).

\end{acknowledgments}

\end{document}